\documentclass{article}

\usepackage{PRIMEarxiv}

\usepackage[utf8]{inputenc} 
\usepackage[T1]{fontenc}    
\usepackage{hyperref}       
\usepackage{url}            
\usepackage{booktabs}       
\usepackage{amsfonts}       
\usepackage{nicefrac}       
\usepackage{microtype}      
\usepackage{lipsum}
\usepackage{fancyhdr}       
\usepackage{graphicx}       
\graphicspath{{media/}}     
\usepackage{amsmath}
\title{Highly efficient non-rigid registration in k-space with application to cardiac Magnetic Resonance Imaging}
\author{
Aya Ghoul\textsuperscript{1}, Kerstin Hammernik\textsuperscript{2}, Andreas Lingg\textsuperscript{3}, Patrick Krumm\textsuperscript{3}, Daniel Rueckert\textsuperscript{2,4,5}, \AND Sergios Gatidis\textsuperscript{1,6}, Thomas Küstner\textsuperscript{1}
}

\begin{document}
\maketitle
\vspace{-1.0cm}
\noindent\footnotesize \textsuperscript{1}Medical Image and Data Analysis (MIDAS.lab), Department of Diagnostic and Interventional Radiology,  University Hospital of Tuebingen, Tuebingen, Germany,  \footnotesize\textsuperscript{2}School of Computation, Information and Technology, Technical University of Munich, Munich, Germany, \footnotesize \textsuperscript{3}Department of Diagnostic and Interventional Radiology, University Hospital of Tuebingen, Tuebingen, Germany, \footnotesize\textsuperscript{4}Klinikum Rechts der Isar, Technical University of Munich, Munich, Germany, \footnotesize\textsuperscript{5}Department of Computing, Imperial College London, London, United Kingdom, 
\footnotesize\textsuperscript{6}Department of Radiology, Stanford University, Stanford, California, USA
\vspace{0.5cm}

\begin{abstract}
In Magnetic Resonance Imaging (MRI), high temporal-resolved motion can be useful for image acquisition and reconstruction, MR-guided radiotherapy, dynamic contrast-enhancement, flow and perfusion imaging, and functional assessment of motion patterns in cardiovascular, abdominal, peristaltic, fetal, or musculoskeletal imaging. Conventionally, these motion estimates are derived through image-based registration, a particularly challenging task for complex motion patterns and high dynamic resolution. The accelerated scans in such applications result in imaging artifacts that compromise the motion estimation. In this work, we propose a novel self-supervised deep learning-based framework, dubbed the Local-All Pass Attention Network (LAPANet), for non-rigid motion estimation directly from the acquired accelerated Fourier space, i.e. k-space. The proposed approach models non-rigid motion as the cumulative sum of local translational displacements, following the Local All-Pass (LAP) registration technique. LAPANet was evaluated on cardiac motion estimation across various sampling trajectories and acceleration rates. Our results demonstrate superior accuracy compared to prior conventional and deep learning-based registration methods, accommodating as few as 2 lines/frame in a Cartesian trajectory and 3 spokes/frame in a non-Cartesian trajectory. The achieved high temporal resolution (less than 5 ms) for non-rigid motion opens new avenues for motion detection, tracking and correction in dynamic and real-time MRI applications.
\end{abstract}


\section{Introduction}\label{sec0}
Magnetic Resonance Imaging (MRI) provides detailed cross-sectional images and tissue characterization without ionizing radiation. Nevertheless, patient movement during acquisition can significantly compromise the quality and diagnostic value of MR scans \cite{slipsager2020quantifying, ling2012head}. The sequentially acquired k-space lines may be scanned in different motion states resulting in ghosting, blurring, ringing, and slice misalignments \cite{zaitsev2015motion}, as a consequence of the spatial encoding in MR physics. Motion contributes to more complex and time-consuming clinical workflows, demanding individualized adjustments and operator expertise to mitigate artifacts. Nearly $20\%$ of clinical MRI exams involve repeated sequences due to motion-degraded images, resulting in an annual cost of about $\$115,000$ per scanner \cite{andre2015toward}.\\
Conventional motion correction techniques, such as triggering, breath-holding and navigator-gating \cite{bluemke1997segmented, vincenti2014compressed, welch2002spherical} often rely on external motion sensors \cite{maclaren2013prospective, falcao2022pilot} or patient cooperation, which can hinder scan efficiency and image quality, particularly in the presence of irregular or unpredictable motion \cite{abi2007cardiac, song2011evaluation}. While segmented sequences \cite{runge1984respiratory, bailes1985respiratory}, fast acquisition schemes \cite{pruessmann1999sense, lustig2007sparse} and motion-resistant sampling trajectories \cite{pipe1999motion, winkelmann2006optimal} have been proposed to address motion artifacts explicitly, they often lack the spatio-temporal resolution required for real-time monitoring of dynamic physiological processes \cite{setser2000quantification, adluru2007temporally}. Existing real-time MRI techniques target image reconstruction, leading to slower scan and processing times. Acquiring sufficient k-space data to achieve adequate spatial resolution can significantly increase imaging time, limiting the attainable temporal resolution \cite{uecker2010real}. These approaches can also suffer from non-convex optimization challenges \cite{yang2016sparse}, geometric distortions, or signal loss due to their sensitivity to off-resonance effects of the magnetic field \cite{ordidge1982real}. Alternatively, the acquisition approach can be tailored to capture motion dynamics independent of reconstruction \cite{huttinga2021real}. While such targeted approaches are effective for streamlining clinical applications, their generalizability may be limited due to application-specific method crafting.\\
Various applications can benefit from highly temporal resolved motion information. In reconstruction, motion can be used to track \cite{marxen2009correcting, herbst2012prospective}, reject motion-corrupted data \cite{cordero2018three, kustner2019retrospective}, or leverage temporal redundancy \cite{cruz2017highly, bustin20203d} to improve image resolution or shorten the acquisition time. In MR-guided radiotherapy, real-time motion tracking enables precise dose delivery to tumors while minimizing radiation exposure to healthy tissues \cite{huttinga2021real, shao2022real}. Moreover, aligning inter-modality and/or intra-modality scans facilitates information fusion from complementary scans. This can be utilized to analyze lesions in dynamic contrast-enhanced MRI \cite{filipovic2011motion}, to compensate inter-frame misalignments in perfusion maps \cite{scannell2019robust}, and to plan and perform image-guided interventions \cite{chuter2017use, risholm2011multimodal}. Real-time motion estimation is particularly valuable in identifying irregular motion patterns, including those encountered in non-cooperative patients \cite{ong2020extreme}, arrhythmia \cite{ferreira2013cardiovascular}, fetal MRI \cite{singh2020deep} and flow imaging \cite{markl2016advanced}. Motion behavior can also be leveraged for functional assessment in various applications, such as cardiovascular \cite{suinesiaputra2009automated, phatak2009strain}, abdominal \cite{schwizer2006magnetic}, peristaltic \cite{heye2012evaluation}, upper airway structures \cite{chen2019intermittently} or musculoskeletal \cite{borotikar2017dynamic} imaging.\\
Despite the potential benefits of motion information across various applications, its broader use is hindered by the challenges of estimating spatially dense motion displacements in non-rigid moving body structures, especially from highly accelerated data or for high temporal resolutions. Conventional registration algorithms \cite{cruz2017highly, bustin20203d} and statistical models \cite{phatak2009strain, suinesiaputra2009automated}, while effective, often suffer from lengthy computation times and require subject-specific adjustments. Deep learning-based registration models have addressed these limitations by reducing processing times and achieving state-of-the-art performance \cite{balakrishnan2019voxelmorph, usman2020retrospective, qi2020non, ghoul2024attention}. However, most of these methods rely on image-space formalism which requires adequate image quality to perform accurate image registration. 
\begin{figure*}[!t]
\includegraphics[width=\linewidth]{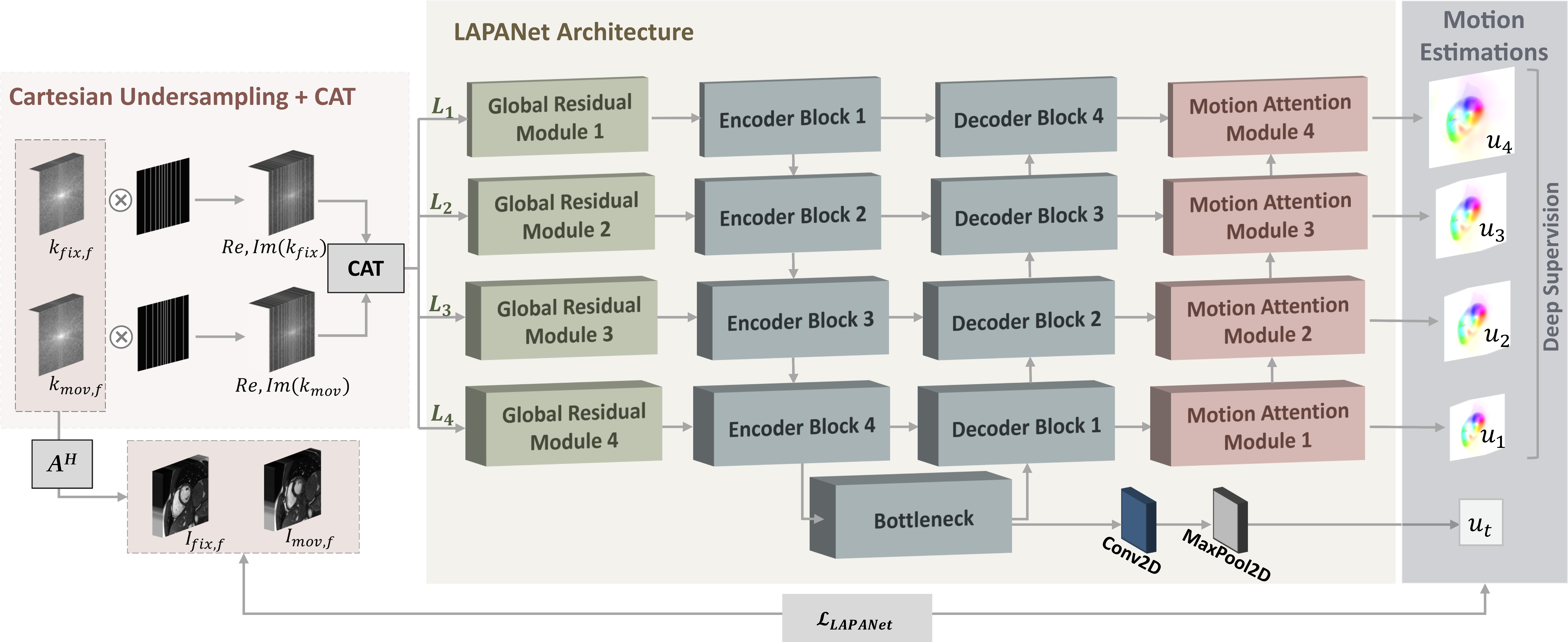}
\label{architecture}
\caption{LAPANet architecture for non-rigid registration in k-space with exemplary Cartesian undersampling. The accelerated coil-resolved fixed ($k_{fix}$) and moving ($k_{mov}$) k-spaces are obtained by undersampling the fully sampled fixed ($k_{fix,f}$) and moving ($k_{mov,f}$) k-spaces during training. $A^{H}$ refers to the multi-coil backward operation, which is used for loss calculation $\mathcal{L}_{LAPANet}$. The real and imaginary parts of the coil-resolved $k_{fix}$ and $k_{mov}$ are concatenated (CAT) to create the input. \textit{Global Residual Modules} (Fig.~\ref{Global_Residual_Module}) extract multi-scale k-space features at $4$ levels, denoted as $L_i$. The \textit{Encoder and Decoder Blocks} (Fig.~\ref{Encoding_Decoding_Block}) extract local and global representations relying on transformer modules and convolutional operations with varying strides. The \textit{Motion Attention Modules} (Fig.~\ref{Motion_Attention_Module}) refine the motion estimation $u_i$ of the current level. We also learn the global translational motion $u_t$ that aligns the moving image to the fixed image.}
\end{figure*}
This in turn limits their temporal resolution, hindering the possibility of accelerating imaging for motion estimation. As a workaround, motion estimates were obtained from low-resolution scans \cite{hansen2012retrospective}. However, these techniques are limited to periodic low-frequency sampling trajectories \cite{huttinga2021real, terpstra2020deep}, require motion regularization \cite{olausson2023time}, and remain subject to the propagation of residual aliasing artifacts into the registration outcomes.\\
K-space-based approaches have been explored to circumvent the limitations of image-domain registration for accelerated acquisitions. Most studies have focused on rigid motion \cite{hossbach2023deep} or required additional previous scans \cite{shao2022real}. Recently, we introduced the LAPNet method for non-rigid registration in k-space \cite{kustner2021lapnet}. LAPNet models non-rigid motion as the cumulative sum of local translations, adhering to the principle of the Local-All Pass (LAP) technique \cite{gilliam20163d}. These translations correspond to phase shifts in k-space, which LAPNet learns on a patch-wise level. Despite promising results, several challenges remain. First, the supervised training using simulated motion causes performance shortcomings when applied to real-world data. Second, patchwise processing limits information sharing between regions and requires empirical determination of the processing window size based on the field of view and motion type. Third, LAPNet has only been tested for acceleration factors up to $R=30$, which may be insufficient for applications demanding higher temporal resolution.\\
Here, we introduce the Local-All Pass Attention Network (LAPANet), a novel self-supervised deep learning-based framework to perform non-rigid registration directly in k-space. Extending our previous LAP-based approach to whole-image processing, LAPANet achieves high-precision dense motion estimates at unprecedented accelerations. The key contributions can be summarized as follows: (1) LAPANet demonstrates accurate motion estimation while achieving a high temporal resolution in the millisecond range. Reliable registration is attained with as few as $2$ lines per temporal frame ($R=78$,  4.24 ms) in a Cartesian trajectory and $3$ spokes per temporal frame ($R=104$,  4.99 ms) in a non-Cartesian radial sampling. Other conventional and deep learning-based methods fail to model or predict motion correctly when confronted with such high accelerations. (2) Our network learns from the full-sized k-space over multi-resolution levels to integrate short- and long-range features of multi-coil information from the available accelerated data adaptively. This holistic approach offers a more comprehensive understanding compared to the localized processing of LAPNet. (3) In contrast to existing k-space-based registrations, the proposed network is designed to uniquely solve the registration task independent of application-specific requirements, downstream tasks, preceding scans, simulations, and ground truth motion fields. This reduces computational demands and minimizes potential biases, promoting improved generalization.\\
We demonstrate the effectiveness of our approach for cardiac motion estimation in a cohort comprising $96$ patients with suspected cardiovascular diseases and $38$ healthy subjects in a retrospective undersampling setting. We conduct an interpretability analysis of LAPANet to examine how different regions within the input k-space contribute to the prediction. Our results indicate a significant increase in temporal resolution for motion estimation across different sampling trajectories and the potential for reliable real-time performance.\\
The remainder of this paper is structured as follows. In Section \ref{sec2}, a detailed description of our novel method and experimental setting is provided. Section \ref{sec3} presents our results. The discussion and conclusion are drawn in Section \ref{sec4} and Section \ref{sec5}, respectively.
\section{Methods}\label{sec2}
\begin{figure*}[!t]
\centering
\includegraphics[width=\textwidth]{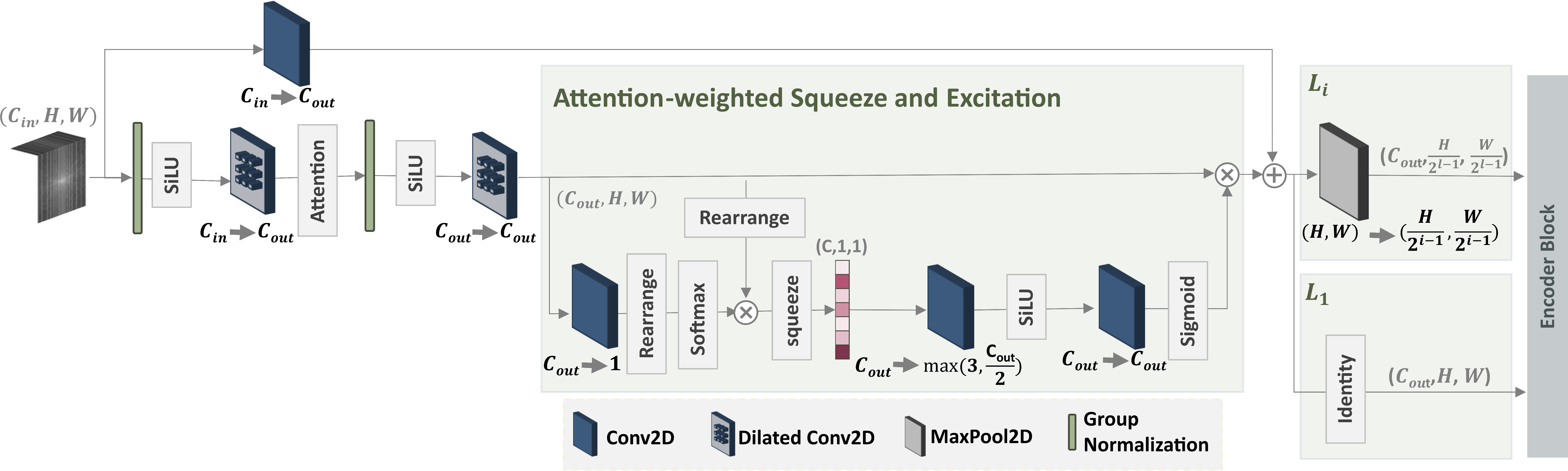}
\caption{Illustration of the \textit{Global Residual Module}. The full-sized stacked k-spaces of shape $(C_{in}, H, W)$ are processed through a residual block connection structure to output a feature map of shape $(C_{out}, H, W)$ for the first network level $L_1$ and  $(C_{out}, H/2^{(i-1)}, W/2^{(i-1)})$ for the other levels $L_i$. The cross-layer connection involves convolutional mapping. The main connection incorporates a self-attention module, replacing traditional linear projection with depthwise convolutions to preserve spatial context. Channel-wise convolution is applied for coil weighting and information storage. Next, the \textit{Attention-weighted Squeeze and Excitation Block} recalibrates channel-wise responses by squeezing the global spatial information into a channel descriptor. An attention mechanism learns subsequently channel-specific weights, used to excite the original feature maps to focus dynamically on important channels.}\label{Global_Residual_Module}
\end{figure*}
\begin{figure*}[h!]
\centering
\includegraphics[width=\textwidth]{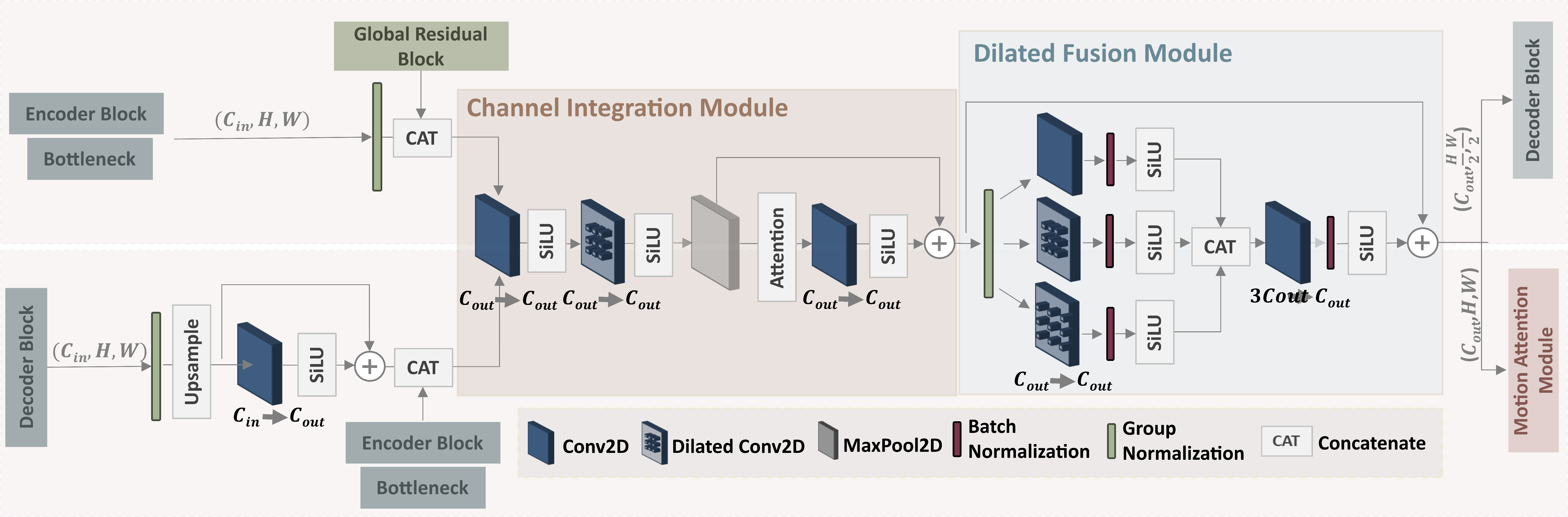}
\caption{Illustration of the encoding and decoding blocks. Encoding combines the input of shape $(C_{in}, H, W)$  with the \textit{Global Residual Module} output. Decoding involves first upsampling using the nearest neighbor while integrating skip connection features. Both encoding and decoding Blocks comprise subsequently a \textit{Channel Integration Module}, followed by a \textit{Dilated Fusion Module} and output feature maps of shape $(C_{out}, H/2, W/2)$ and $(C_{out}, 2H, 2W)$ respectively. Sigmoid-weighted Linear Unit (SiLU) activations are utilized throughout. Additionally, a self-attention mechanism is employed by projecting the features into Queries, Keys, and Values using depthwise convolutions to preserve spatial context and leverage correlations across the channel dimension}
\label{Encoding_Decoding_Block}
\end{figure*}
\subsection{The Local-All Pass (LAP) principle}
We assume local brightness consistency, where intensity remains constant when transitioning from a moving $I_{mov}$ to a fixed $I_{fix}$ image. These images can be connected locally through a translation $\underline{u}_{\text{trans}}(\underline{x})$ at the pixel position $\underline{x}$, which corresponds to a shifting property in the Fourier domain (k-space). This results in a multiplication of the moving k-space $k_{mov}$ with a linear phase to obtain the fixed k-space $ k_{fix}$ at all k-space locations $\underline{k}$. Hence, the corresponding translation can be expressed equivalently in the Fourier domain as:
\begin{equation}
I_{fix}(\underline{x}) = I_{mov}(\underline{x}-\underline{u}_{\text{trans}}) \rightleftharpoons k_{fix}(\underline{k}) = k_{mov}(\underline{k}) e^{-\text{j}\underline{u}_{\text{trans}}^\text{T}\underline{k}} \label{eq0}
\end{equation}
where,
\begin{align} 
H(\underline{k}) = e^{-j\underline{u}_{\text{trans}}^\text{T}\underline{k}}  \label{eq00}
\end{align}
is an all-pass filtering operation in the Fourier domain \cite{gilliam20163d}. By extending this global translation to a localized translation in the image domain, we can model non-rigid motion as a superposition of local translational displacements within small windows. In the Fourier domain (k-space), the analogous process results in a superposition of phase shifts, or filtering with all-pass filters, adhering to the core principle of the Local-All Pass (LAP) algorithm. The tapered k-spaces corresponding to local windowing in the image domain can be generated using a convolution with a phase-modulated tapering function \cite{kustner2021lapnet}. 
\subsection{LAPANet Architecture}
We introduce the Local All-Pass Attention Network (LAPANet), illustrated in Fig.~\ref{architecture}, a deep learning-based framework for approximating the Local-All Pass filters from the accelerated k-space data to extract non-rigid motion information. LAPANet operates on four levels \{$L_1, L_2, L_3, L_4$\} to enable multi-scale processing. The input k-spaces size is selected to be $160\times160$, leading to a feature map size of $5\times5$ in the bottleneck block. We stack the real and imaginary components of the coil-resolved fixed and moving k-spaces to create a real-valued 2D input. Then, we apply a circular shift operation to relocate the low-frequency components from the center of the spectrum to the edges while shifting the high-frequency components to the center, i.e. inverse zero-frequency shift operation \cite{khare2023operational}. This operation has proven beneficial in capturing intricate local features while expediting convergence.\\
The \textit{Global Residual Modules} establish multi-resolution k-spaces by partitioning the combined input k-spaces into local windows of varying sizes. This operation mimics the phase-modulated tapering functions \cite{kustner2021lapnet} in the Fourier domain, corresponding to extracting patches in the image domain. At each level, the processing window size is reduced by a factor of $2$ to prepare for a coarse-to-fine estimation paradigm. The built partitions of k-spaces serve as an additional input to the encoding blocks. Transferring the computation to a coarser scale also optimizes memory usage without boundaries ambiguity and loss of fine details, unlike downscaling in the image domain.\\
\begin{figure*}[h!]
\centering
\includegraphics[width=0.9\textwidth]{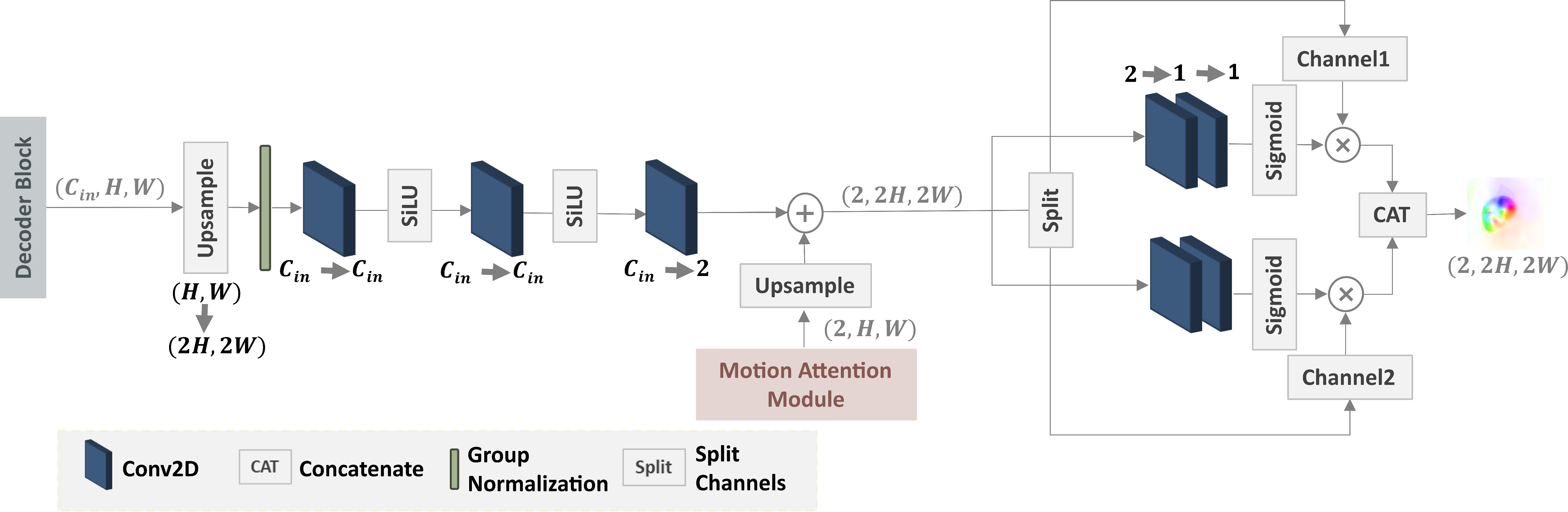}
\caption{Illustration of the \textit{Motion Attention Module}. This module involves upsampling using the bilinear interpolation by a factor of 2, followed by feature encoding from $(C_{in}, 2H, 2W)$ to $(2, 2H, 2W)$. Then, a refining operation improves the current decoder estimation by combining current and previous motion estimates. Separate attention masks for the two channels (Channel 1, Channel 2) corresponding to the spatial dimensions of the image are learned. The motion estimation channels are then weighted individually using the learned attention masks for dimension-specific fine-tuning. Finally, these weighted channels are concatenated (CAT) to obtain the motion estimation at the current level.} \label{Motion_Attention_Module}
\end{figure*}
The subsequent building blocks of LAPANet include encoding and decoding paths with a skip connection topology. Each encoding and decoding block sequentially incorporates a \textit{Channel Integration Module}, followed by a \textit{Dilated Fusion Module}. Finally, the decoding path is enhanced by the \textit{Motion Attention Modules}, which integrate features from preceding levels to refine the current motion estimation and output dense motion estimates  \{$u_1, u_2, u_3, u_4$\} with different resolutions. Moreover, we learn the global translational motion $u_t$, which maps the moving image to the fixed image. This encourages the encoding blocks to learn more meaningful features for the motion estimation task. We establish the mapping from the bottleneck block using a convolutional operation with $1\times1$ kernel followed by a max pooling with $5\times5$ kernel. In the following sections, we elaborate on the implementation details of the key building blocks of our method. We refer to convolutional layers with  $3\times3$ kernel, 1-stride and a dilation $k$ as $Conv_{d_k}$.
\subsubsection{Global Residual Module}
The \textit{Global Residual Module} (GRM), depicted in Fig.~\ref{Global_Residual_Module}, has a residual block connection structure and takes as input the full-sized stacked k-spaces. The cross-layer connection contains a convolutional mapping with one convolutional layer to extract the wanted number of feature maps, that starts at $4$ and is increased to $16$, $32$, and $128$ in the following levels. The main connection contains a self-attention module between two convolutional operations that include a batch normalization layer, followed by a Sigmoid-weighted Linear Unit (SiLU) activation \cite{elfwing2018sigmoid} and a $Conv_{d_2}$ layer. The self-attention block transforms the extracted feature maps into Keys, Queries, and Values through depthwise convolutions so that each channel is convolved with its own set of filters. Unlike traditional linear projection used in transformers, this strategy preserves spatial context without positional encoding and extracts information from the channel dimension where coil weighting and information across the input k-spaces are stored.\\
The extracted maps are readjusted dynamically using the subsequent \textit{Attention-weighted Squeeze and Excitation Block}. We leverage squeeze-and-excitation units \cite{hu2018squeeze} to retain the most valuable features and discard less relevant ones. We address inter-channel relationships to learn specific constant weights for each feature channel of every input. The feature maps of dimensions $H \times W \times C$ undergo projection using a $Conv_{d_1}$ operation, followed by a softmax operation. The outcome of this projection and the original feature maps are then reshaped into a flattened representation with dimensions $H W \times 1$ and $H W \times C$. Subsequently, we perform element-wise multiplication of both maps and compress the result through a global average pooling operation. The outcome is a feature descriptor with dimensions ($C\times1\times1$) that encapsulates attention weights for the channel dimension. The following excitation process involves sequential $Conv_{d_1}$-SiLU activation-$Conv_{d_1}$-Sigmoid activation operations. The resultant channel-wise weights are multiplied by the initial features to generate the attention-weighted feature map. Finally, we use max pooling operations to downscale the feature dimensions for levels $L_2, L_3$, and $L_4$ using $2 \times 2$, $4 \times 4$, and $8 \times 8$ kernels with $2$, $4$, and $8$ strides respectively and, hence, create the multi-scale inputs for the subsequent encoding units.   
\begin{figure*}[h]
\centering
\includegraphics[width=\textwidth]{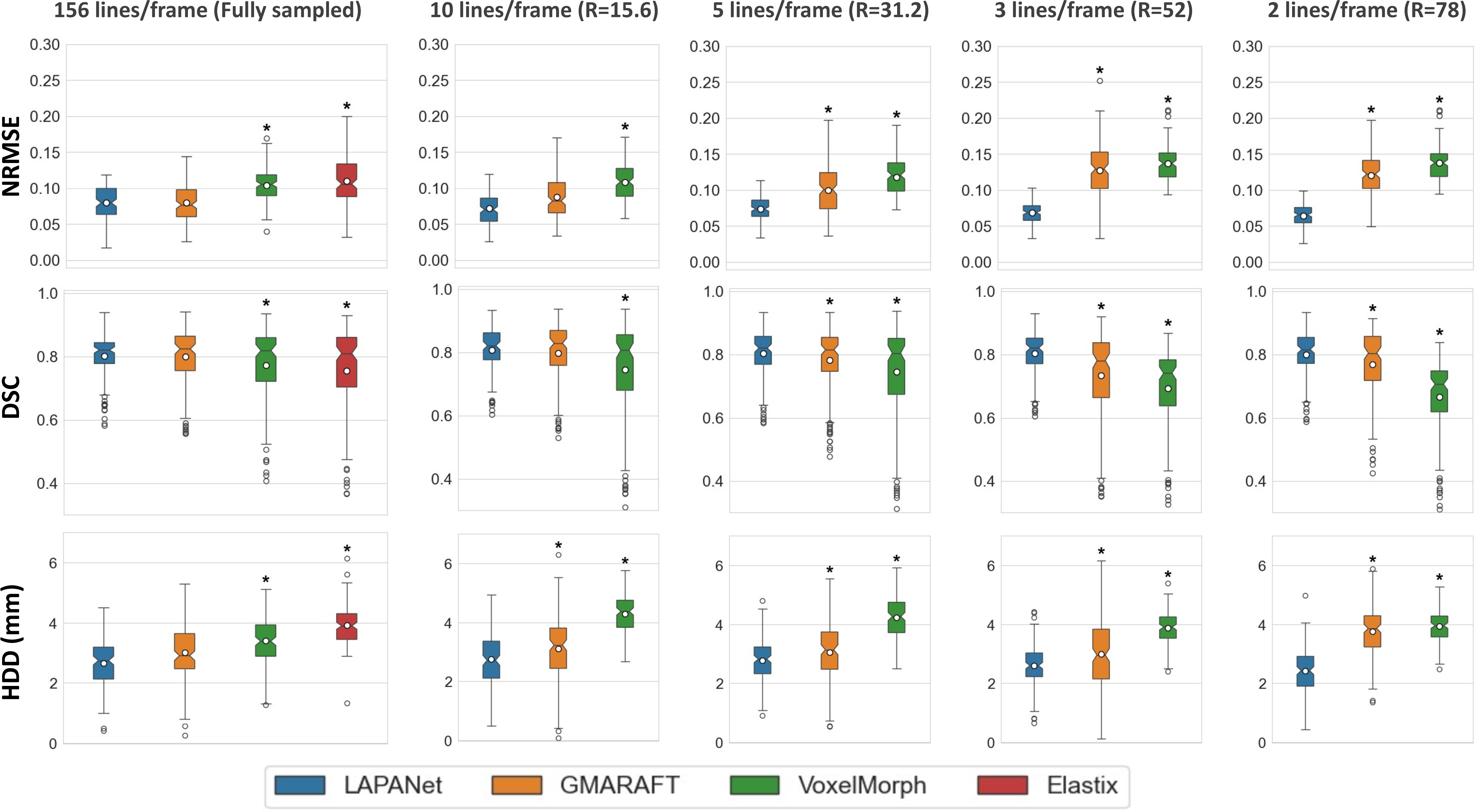}
\caption{Boxplots of normalized root-mean-square error (NRMSE), Dice scores (DSC), and Hausdorff Distances (HDD) after registration with the predicted motion from the proposed LAPANet in comparison to GMA-RAFT, VoxelMorph, and Elastix. Metrics are shown for motion estimation in the fully sampled case and at different accelerations using a Cartesian VISTA sampling on the test dataset. The symbol $*$ denotes a statistically significant difference, indicated by $P<0.05$ when compared to LAPANet. LAPANet outperformed other image-based methods, maintaining superior scores across different accelerations. Alternative methods yielded continuous performance degradation with increased accelerations. Elastix was not able to predict any motion in these highly accelerated cases and was omitted.}
\label{quantitative_VISTA}
\end{figure*}
\subsubsection{Encoding and Decoding Blocks}
The encoding and decoding blocks are illustrated in Fig.~\ref{Encoding_Decoding_Block}. The input undergoes initially a group normalization. In the encoding pathway, the result is combined with the output from the corresponding \textit{Global Residual Module} and subsequently fed into the succeeding modules. In the decoder blocks, an upsampling by a factor of $2$ is applied, followed by a residual unit comprising a $Conv_{d_1}$ operation and a SiLU activation. The resultant output is then stacked with the skip connection features.\\
The next module, the \textit{Channel Integration Module} (CIM), starts with a sequence of $Conv_{d_1}$-SiLU activation-$Conv_{d_1}$-SiLU activation-$Conv_{d_2}$-SiLU activation operations. A max pooling with a $2 \times 2$ kernel is then added to the encoder units. Next, a residual unit with one self-attention block-$Conv_{d_1}$-SiLU activation in the main connection is used. The attention module is identical to the attention operation used in the \textit{Global Residual Module}. The number of convolutional filters begins at $16$ at level $L_1$ and progressively increases to $32$, $64$, and $192$ as the network processes the following levels, reaching a maximum of $384$ filters in the bottleneck. The decoding path mirrors the encoder's architecture, with a symmetric decrease in filter count.\\
For more spatial context interpretation, different dilated convolutions are applied in parallel in the \textit{Dilated Fusion Module} (DFM) and combined in a multi-branch convolution module within a residual structure. The residual connection has a group normalization layer followed by three $Conv_{d_k}$-SiLU activation-Batch normalization operations with dilations of $k=\{1,2,4\}$. The stack of the dilated convolutions is subsequently encoded with $Conv_{d_1}$-SiLU activation-Batch normalization. 

\subsubsection{Motion Attention Module}
The \textit{Motion Attention Module} (MAM), shown in Fig.~\ref{Motion_Attention_Module}, aggregates multi-resolution information at multiple concatenation points to perform upsampling and provide dimension-specific fine-tuning. The output of each decoder block is first upsampled by a factor of $2$. Then, encoding is achieved through three sequential $Conv_{d_1}$-SiLU activation operations to find an initial motion estimation corresponding to the scale at hand. This estimation is stacked with the upsampled estimation from the previous \textit{Motion Attention Module} to fuse features across different resolutions. Next, we learn two pixel-wise weighting maps with values ranging in $[0,1]$ for the two spatial channels of the motion estimation. The two maps are built using a sequence of separate $Conv_{d_1}$-$Conv_{d_1}$-Sigmoid activation operations. Based on the learned attention masks, the motion estimation channels are individually adjusted and concatenated to create the final motion estimation for the current level.

\begin{figure*}[h]
\centering
\includegraphics[width=\textwidth]{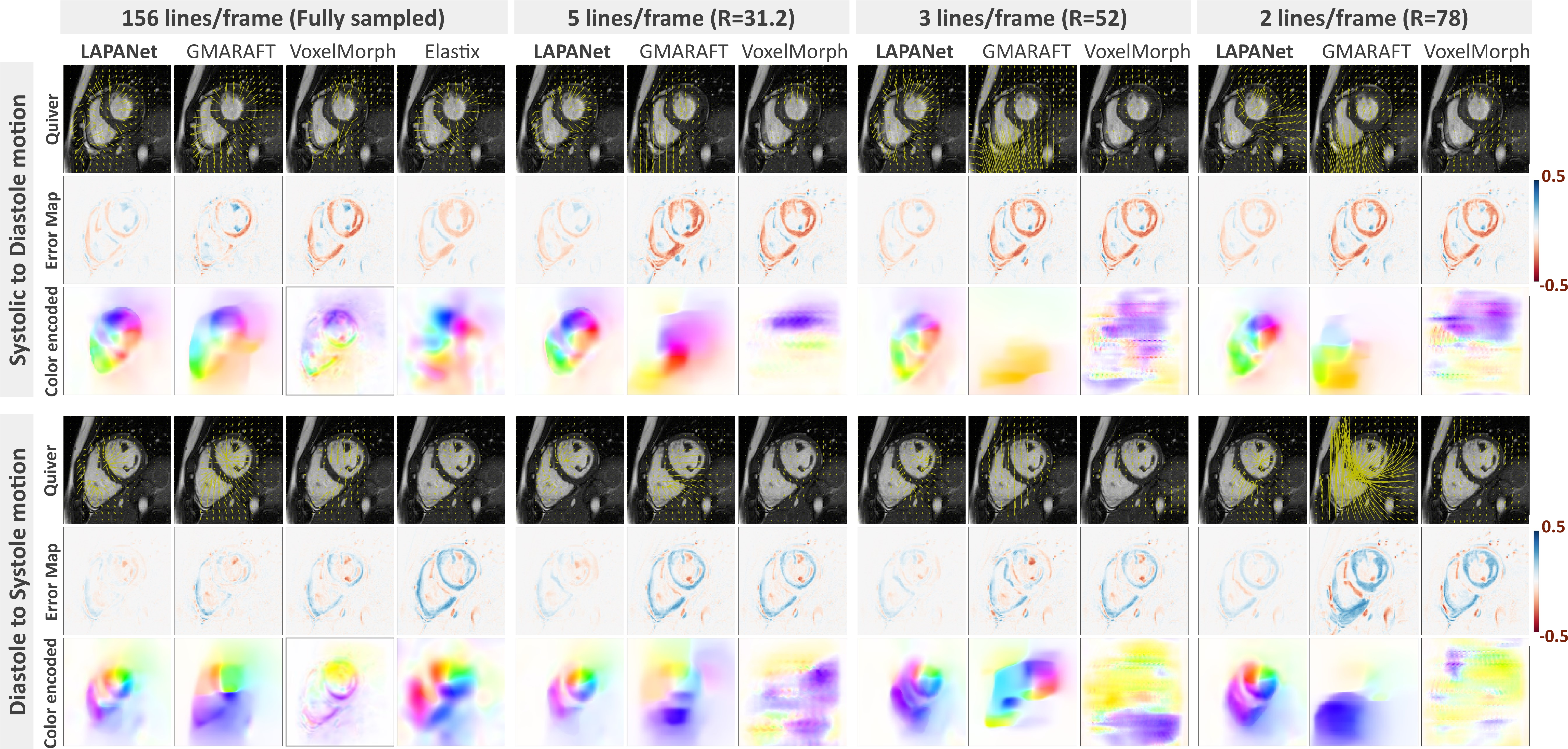}
\caption{Motion estimation between i) systolic to diastolic (top rows) and ii) diastolic to systolic (bottom rows) cardiac cine frames of a patient with suspected right ventricular cardiomyopathy. Accelerated data is obtained with retrospective undersampling using the VISTA mask for the fully sampled case and at three accelerations ($R=31.2$, $R=52$, and $R=78$).  Motion estimation is shown for the proposed LAPANet compared to GMA-RAFT, VoxelMorph, and Elastix. Results are represented with quiver plots overlaid on the fully sampled moving image (first row), error maps between the fully sampled fixed and warped moving image using the estimated motion (second row), and color-encoded \cite{baker2011database} motion estimates (third row).} \label{qualitative_VISTA}
\end{figure*}

\subsection{Database} 
We used LAPANet to estimate cardiac motion from pairs of 2D cine data for fully sampled and retrospectively accelerated cases. Multi-slice short axis 2D bSSFP cine scans were acquired in-house on a $1.5$T MRI (MAGNETOM Aera, Siemens Healthineers, Erlangen, Germany). The dataset contains $134$ subjects, including $96$ patients ($45\pm17$ years, $24$ female) with suspected cardiovascular diseases and $38$ healthy subjects ($32\pm6$ years, $14$ female). 2D cine was acquired over $8$ breath-holds of $15$ s duration each. Other parameters included: TE=$1.06$ ms, TR=$2.12$ ms, resolution= $1.9$$\times$$1.9$$ \text{mm}^2$ with a matrix size in the range of $176 \times 132$ to $192 \times 180$, slice thickness=$8$ mm, temporal resolution in the range of $18$ to $80$ ms, $25$ temporal phases, flip angle=$52^\circ$ and bandwidth=$915$ Hz/px.\\ 
Ground truth segmentations of the left and right ventricles were initially generated with the Segment software \cite{heiberg2010design} and then corrected manually. Coil sensitivity maps were determined from the acquired auto-calibration signal data following \cite{uecker2014espirit} and then coil-compressed with singular value decomposition (SVD) to $n_c=16$. All subjects were prospectively recruited and gave written consent for participation in the trial and publication of their imaging data and results. The local ethics committee approved the study (426/2021BO1).\\
The network was trained on $96$ subjects ($25$ healthy subjects and $72$ patients), validated on another $12$ subjects ($4$ healthy subjects and $8$ patients) and tested on $25$ held-out subjects ($9$ healthy subjects and $16$ patients). Cartesian undersampling masks followed the variable density incoherent spatio-temporal acquisition (VISTA) \cite{ahmad2015variable}, while a golden angle radial sampling was used for the non-Cartesian sampling \cite{winkelmann2006optimal}. The acceleration factor $R$ was determined by dividing the frequency encoding lines or spokes per frame on the undersampled grid by the lines or spokes on the fully sampled grid. During training, accelerations were selected in the range of $R=1$ (fully sampled), to $R=78$ for the Cartesian trajectory and $R=104$ for radial sampling, corresponding to $2$ and $3$ remaining lines or spokes per temporal frame, respectively.\\ 
All possible pairs of fixed and moving frames throughout the cardiac cycle were included. We created a total of $647,500$ training image pairs and $171,875$ testing image pairs. The fixed and moving input images were coil-resolved, central cropped, and zero-padded to $160 \times 160$ pixels, and divided by their maximum intensity value for normalization. Then, a Fourier operation mapped the images to k-space to form the input for training. During inference, we performed the registration on arbitrary-sized inputs by recovering the original full size from the prediction using zero-padding. 

\subsection{Training Loss and Implementation Details}
We used the coil-resolved fixed $I_{fix,f}$ and moving $I_{mov,f}$ images for loss computations. These were obtained with the multi-coil backward operation $A^H$, which includes an inverse Fourier transform, applied on the fully sampled input fixed and moving k-spaces. Self-supervision and deep supervision were combined for the training. We used the photometric loss $\mathcal{L}_{photo}$ between the fixed image $I_{fix,f}$ and the warped image, obtained by transforming the moving image $I_{mov,f}$ with the upscaled estimated motion $u_i$ obtained at the scaling level $L_i$ using a bilinear interpolation operation $T$. We incorporate a bounding box $\phi$, defined by a fixed $10$-pixel margin around segmented left and right ventricles. This emphasizes the heart region and mitigates the impact of peripheral disturbances by preserving values within and nullifying those outside the box:
\begin{equation}
\mathcal{L}_{photo, i} = ||\phi(I_{fix,f} - T(I_{mov,f}, u_i))||_1 \label{eq1}
\end{equation}
A data consistency loss $\mathcal{L}_{\text{K-DC}}$ that operates in k-space was included since the k-space data of the warped images should be consistent with the k-space data of the fixed images if the registration model can generate a high-quality motion estimation from any undersampled coil-resolved k-space. We included only the difference in amplitudes, as at small values in k-space, the gradient derived from the phase exhibited discontinuities or exploded which disturbed the training:
\begin{equation}
\mathcal{L}_{\text{K-DC}, i} = \Big|\Big| ||\mathcal{F}(I_{fix,f})||_2 - ||\mathcal{F}(T(I_{mov,f}, u_i))||_2 \Big|\Big|_1 \label{eq2}
\end{equation}
The smoothness loss $\mathcal{L}_{smooth}$ is added to avoid local discrepancies and is obtained with an anisotropic diffusion regularizer on the spatial gradients of the motion estimate. The Jacobian $\nabla$ is determined by estimating spatial gradients along image directions using forward finite differences to compute the partial derivatives:
\begin{equation}
\mathcal{L}_{smooth, i} = ||\nabla u_i||_1 \label{eq3}
\end{equation}
Additionally, we incorporated the global translational photometric loss $\mathcal{L}_{Tphoto}$ to learn the translation $u_t$ between the fully-sampled fixed and moving images corresponding to the input k-spaces and, thus, encourage the encoder to extract more meaningful representations for the registration task:
\begin{equation}
\mathcal{L}_{Tphoto} = ||\phi(I_{fix,f} - T(I_{mov,f}, u_t))||_1 \label{eq4}
\end{equation}
The total training loss $\mathcal{L}_{total}$ was computed by aggregating all individual loss functions, as follows:
\begin{equation}
\mathcal{L}_{total} =  \alpha\mathcal{L}_{Tphoto} + \sum_{i=1}^{L=4} \mathcal{L}_{photo, i} + \beta\mathcal{L}_{\text{K-DC},i} + \gamma\mathcal{L}_{smooth,i}\label{eq5}
\end{equation}
where the hyper-parameters $\alpha$, $\beta$ and $\gamma$ are set to $0.5$, $0.05$ and $0.01$. These parameters were optimized using random search to minimize the photometric loss $\mathcal{L}_{photo}$ on the validation subjects. We used an AdamW optimizer \cite{loshchilov2017decoupled} (batch size=$32$, learning rate=$1e^{-4}$, weight decay=$1e^{-3}$) with a cosine annealing schedule \cite{loshchilov2016sgdr} for training. The model has $~17.2$ million trainable parameters. We report an average training duration of $~2$h/epoch, i.e. $~120$ h for ~$60$ epochs on $2$ GPUs (NVIDIA V100 GPU) and we observed an average GPU-accelerated estimation time of $~30$ ms per pair of frames.

\begin{figure*}[h]
\centering
\includegraphics[width=\textwidth]{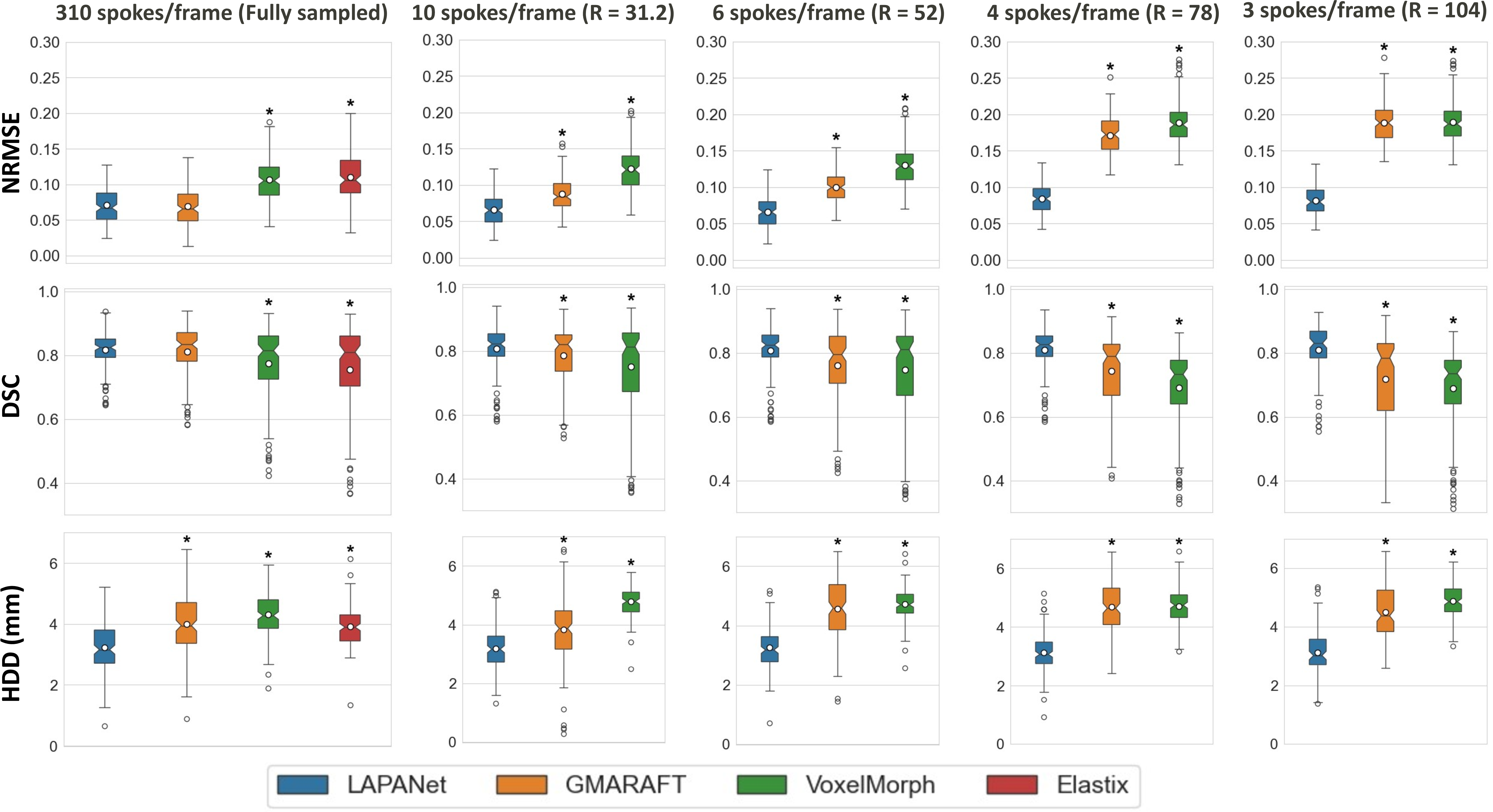}
\caption{Boxplots of normalized root-mean-square error (NRMSE), Dice scores (DSC), and Hausdorff Distances (HDD) after registration with the predicted motion from the proposed LAPANet in comparison to GMA-RAFT, VoxelMorph, and Elastix. Metrics are shown for motion estimation in the fully sampled case and at different accelerations using a non-Cartesian Radial sampling on the test dataset. The symbol $*$ denotes a statistically significant difference, indicated by $P<0.05$ when compared to LAPANet. LAPANet showed improved performance compared to competing image-based registrations across different acceleration factors. Elastix was not able to predict any motion in these highly accelerated cases and was omitted.}
\label{quantitative_Radial}
\end{figure*}

\begin{figure*}[h]
\centering
\includegraphics[width=\textwidth]{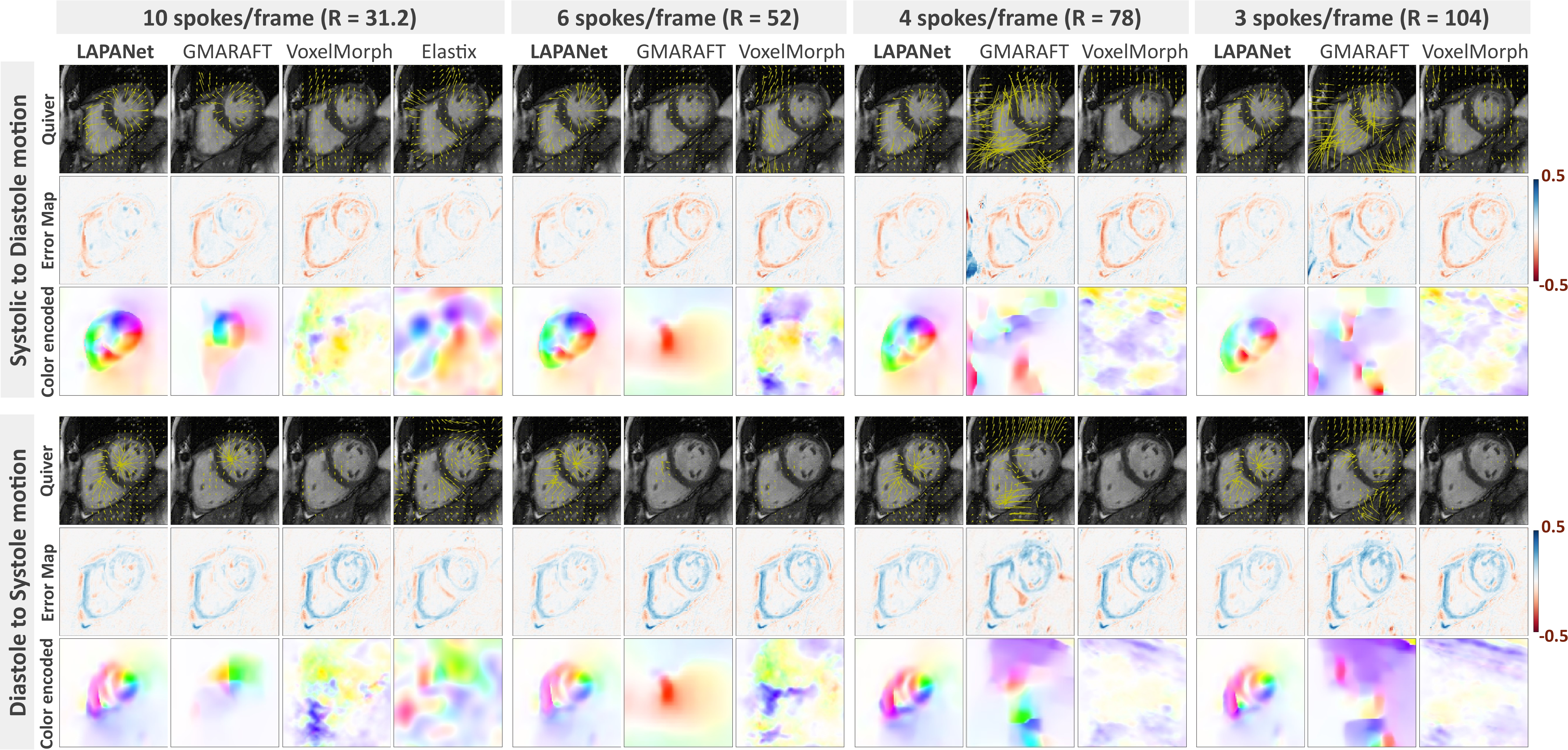}
\caption{Motion estimation between i) systolic to diastolic (top rows) and ii) diastolic to systolic (bottom rows) cardiac cine frames of a healthy subject. Accelerated data is obtained with retrospective undersampling using the radial mask at four accelerations ($R=31.2$, $R=52$, $R=78$, and $R=104$).  Motion estimation is shown for the proposed LAPANet in comparison to GMA-RAFT, VoxelMorph, and Elastix. Results are represented with quiver plots overlaid on the fully sampled moving image (first row), error maps between the fully sampled fixed and warped moving image using the estimated motion (second row), and color-encoded \cite{baker2011database} motion estimates (third row).} \label{qualitative_Radial}
\end{figure*}
\subsection{Experiments}
\subsubsection{Baseline Comparison}
We compared the final motion estimation achieved by LAPANet ($u_4$) against the outcomes obtained by a conventional (Elastix \cite{klein2009elastix}) and two deep learning methods (GMA-RAFT \cite{ghoul2024attention} and VoxelMorph \cite{balakrishnan2019voxelmorph}) on cardiac motion estimation for fully-sampled and accelerated cases using a Cartesian VISTA and a radial sampling. Optimal hyperparameter combinations for all the studied models were determined through a random search with the objective of minimizing the photometric loss $\mathcal{L}_{photo}$ on the validation subjects. Supplementary Table A1 lists the average training time, the inference time, the number of floating-point operations (\#FLOP), and the number of trainable parameters for all studied models. For Elastix, we used $4$ multi-resolution levels and $64$ histogram bins. At every resolution, the optimization undergoes $1000$ iterations guided by the Parzen Window Mutual Information \cite{mattes2001nonrigid} for $2048$ randomly sampled spatial coordinates. Both GMA-RAFT and VoxelMorph were trained with a self-supervised strategy using the hyperparameter settings suggested in \cite{ghoul2024attention}. 

\subsubsection{Evaluation}
We examined the registration performance of each model on two downstream tasks. First, we evaluated the disparity between the transformed images, obtained by warping the fully sampled moving images with the estimated motion and the corresponding fully sampled fixed images using the normalized root-mean-square error (NRMSE). This is useful for motion correction in reconstruction or time-resolved motion identification. Second, we registered the segmentation masks between the most moved frames, i.e. the end-systolic and the end-diastolic, which can be applied later in functional analysis such as left ventricular assessment. We reported the average Dice scores (DSC) and Hausdorff Distances (HDD) for two segmentation masks: the blood pool of the right ventricle and the cavity and the myocardium of the left ventricle.\\
The mean and standard deviation of the quantitative metrics were calculated on all test image pairs. Statistical significance was assessed using the one-way analysis of variance (ANOVA) on the evaluation scores of the competing methods relative to LAPANet results. In addition, we performed the Tukey honestly significant difference (HSD), and Bonferroni/Holm post-hoc corrections for all pairwise group comparisons to investigate LAPANet estimation consistency across different accelerations. A p-value of $P\leq0.05$ was considered statistically significant, adjusted for multiple comparisons depending on the experiment.\\
We further examined how high accelerations impacted the precision and consistency of motion estimation qualitatively by comparing color-coded \cite{baker2011database} motion estimates, quiver plots overlaid on the fully sampled moving image, and the mean square error map between the fully sampled fixed and warped moving image with the motion estimation at the investigated acceleration.

\subsubsection{Ablation Studies}
To evaluate the effectiveness of our architecture and training strategy, we conducted a series of ablation studies using radial sampling. We systematically removed building blocks of our architecture, including the \textit{Global Residual Module}, the encoder and decoder components (the \textit{Dilated Fusion Module} and the \textit{Channel Integration Module}), and the \textit{Motion Attention Module}. Additionally, we experimented with various loss function configurations.  We excluded the global translational photometric loss $\mathcal{L}_{Tphoto}$ and data consistency loss $\mathcal{L}_{\text{K-DC}}$ and calculated the training loss on coil-combined images instead of coil-resolved images. We also explored learning explicit multi-scale mapping information by inserting a phase-modulated tapering operation \cite{kustner2021lapnet} before the \textit{Global Residual Modules} to extract multi-scale k-space patches explicitly before feature extraction. Other ablation experiments examined varying levels of coil compression of the sensitivity maps used to create the input k-spaces. The studied numbers of coils spanned from $1$ (coil-combined) to $24$. 

\subsubsection{Model Interpretability}
We studied the interpretability of LAPANet using the Integrated Gradient (IG) heatmaps \cite{sundararajan2017axiomatic} and the corresponding power spectra for the input k-spaces in Cartesian and radial trajectories. IG allocates saliency attributions by integrating the gradients of the output along a path from a baseline to the input. A series of $100$ intermediate k-spaces were generated by linearly interpolating between all-zero baselines and the input k-spaces. This interpolation involved gradually scaling the input k-spaces and adding them to the baseline, creating a sequence of intermediate k-spaces that smoothly transitioned from the baseline to the input. The output's gradients were then calculated along each pixel of the intermediate k-spaces, and aggregated cumulatively, to visualize the most salient regions of the input. Moreover, we created the mean line profile through the center of the power spectra along the frequency and phase encoding directions, to study the distinctive contributions of the different spatial frequencies over the entire test dataset. We applied a deconvolution with the Point Spread Function of each Cartesian sampling signal to account for the different positions of the remaining frequency encoding lines. The mean Noise Power Spectra (NPS) were then obtained by integrating over all line profiles to reveal the spatial distribution of the noise content in the IG maps. 

\begin{figure*}[h]
\centering
\includegraphics[width=\textwidth]{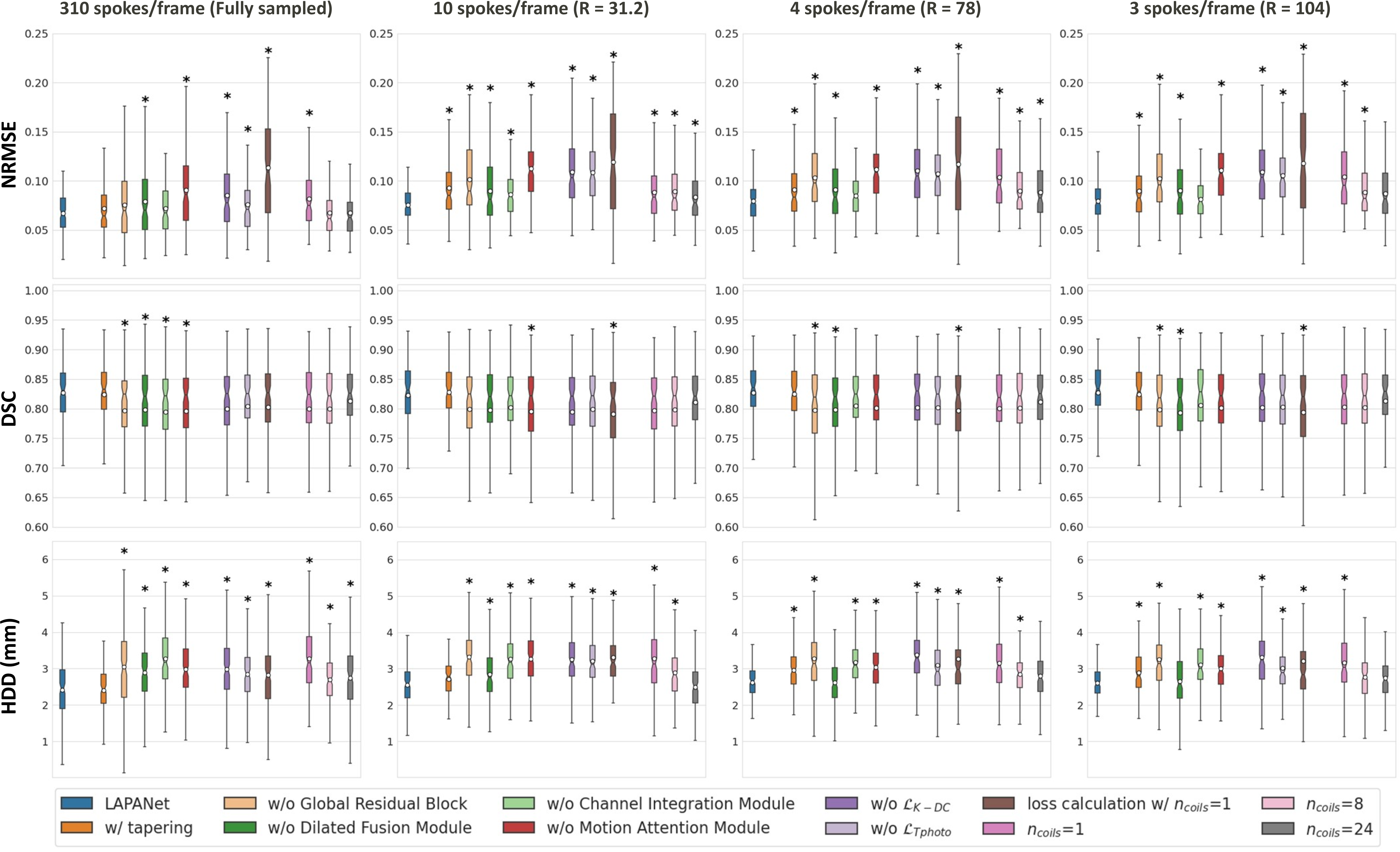}
\caption{Boxplots of normalized root-mean-square error (NRMSE), Dice scores (DSC), and Hausdorff Distances (HDD) for quantitative analysis of the ablation studies in fully sampled, $R=32$, $R=78$, and $R=104$ accelerated data using a radial trajectory (mm). Results are grouped by ablation study along the x-axis including ablations on the building blocks of the network, on the training loss function, and on the coil compression degree of the sensitivity maps used to create the input and for loss computation. $n_c$ indicates the number of remaining coils after compression. Each boxplot represents the distribution of results produced by the ablated network on the test dataset. The symbol $*$ denotes a statistically significant difference, indicated by a $\text{p-value}<0.05$ when compared to the proposed LAPANet.}\label{ablations}
\end{figure*}
\section{Results}\label{sec3}
\subsection{Handling high acceleration across different trajectories}
We compared the performance of LAPANet to other image-based registrations on cardiac motion estimation for the fully sampled case and across different accelerations using the VISTA mask. LAPANet and GMA-RAFT performed better than VoxelMorph and Elastix on all evaluated metrics for the fully sampled case. GMA-RAFT performance began to deteriorate for acceleration rates exceeding $R=31.2$. Conversely, LAPANet consistently maintained performance levels comparable to fully sampled results across all acceleration rates studied, up to $R=78$. This is evident in the boxplots depicted in Fig.\ref{quantitative_VISTA} and the detailed metrics presented in Supplementary Table A2. Our model demonstrated reliable results up to an acceleration factor of $R=78$, equivalent to two remaining encoding lines per temporal frame after acceleration. The other competing image-based registration methods were challenged with such high accelerations, exhibiting a significant increase in registration error as measured by the NRMSE. Both HDD (F-statistic$=0.721$, p-value$=0.801$) and DSC (F-statistic$=0.152$, p-value$=0.962$) remained consistent throughout the investigated accelerations using LAPANet with a mean target registration error $\leq2.8$ mm for initial misalignments of up to $7.2$ mm. On the other hand, image-based approaches exhibited a decline in DSC and HDD with increasing acceleration, coinciding with a degradation in image quality that hinders the visualization of heart structures.\\
Representative motion estimates were investigated in systole and diastole for various accelerations in a patient with suspected right ventricular cardiomyopathy (Fig.~\ref{qualitative_VISTA}). Left ventricular radial motion and right ventricular translational motion were detected by LAPANet across the different accelerations and motion states. Our model achieved the lowest residual error compared to other approaches. As the acceleration increased, a rising trend of misregistered deformations was obtained using the image-based methods. Their performance degraded with increasing acceleration due to intensified aliasing artifacts, as shown in Supplementary Fig. A1. Elastix failed to converge for high accelerations.\\
To further validate the efficacy of our method, we employed a radial trajectory and observed results comparable to those achieved with Cartesian sampling. Image-based approaches demonstrated a decline in performance at higher acceleration rates, whereas LAPANet consistently maintained superior scores up to an acceleration of $R=104$, as illustrated in Fig.~\ref{quantitative_Radial} and Supplementary Table A3. The mean HDD was consistently $\leq3.3$ mm, equivalent to a displacement of $\leq1.7$ pixels, and the mean DSC maintained a value of $0.81$. LAPANet's performance, as indicated by both HDD (F-statistic$=1.105$, p-value$=0.353$) and DSC (F-statistic$=0.043$, p-value$=0.996$), remained stable despite variations in acceleration.\\
Qualitative outcomes of the proposed approach compared to image-based registrations are shown for a healthy subject in Fig.~\ref{qualitative_Radial} using radial undersampling. LAPANet yielded concentrated motion towards the cardiac region against a stationary background and closer alignment with the contraction and relaxation of the heart. GMA-RAFT provided good estimates in fully sampled scenarios. However, severe streaking artifacts, illustrated in Supplementary Fig. A1, at higher acceleration hindered identifying correspondences, eventually obscuring the delineation of heart structures. Similarly, VoxelMorph demonstrated more flow misregistrations with increasing acceleration. Elastix still demonstrated a capability to discern the movement of the left ventricle, albeit with imprecise delineation of the cardiac contours for  $R=31.2$ acceleration. Nevertheless, it failed to yield estimations for all test subjects at such high accelerations and diverged in most cases.

\begin{figure*}[h]
\centering
\includegraphics[width=\textwidth]{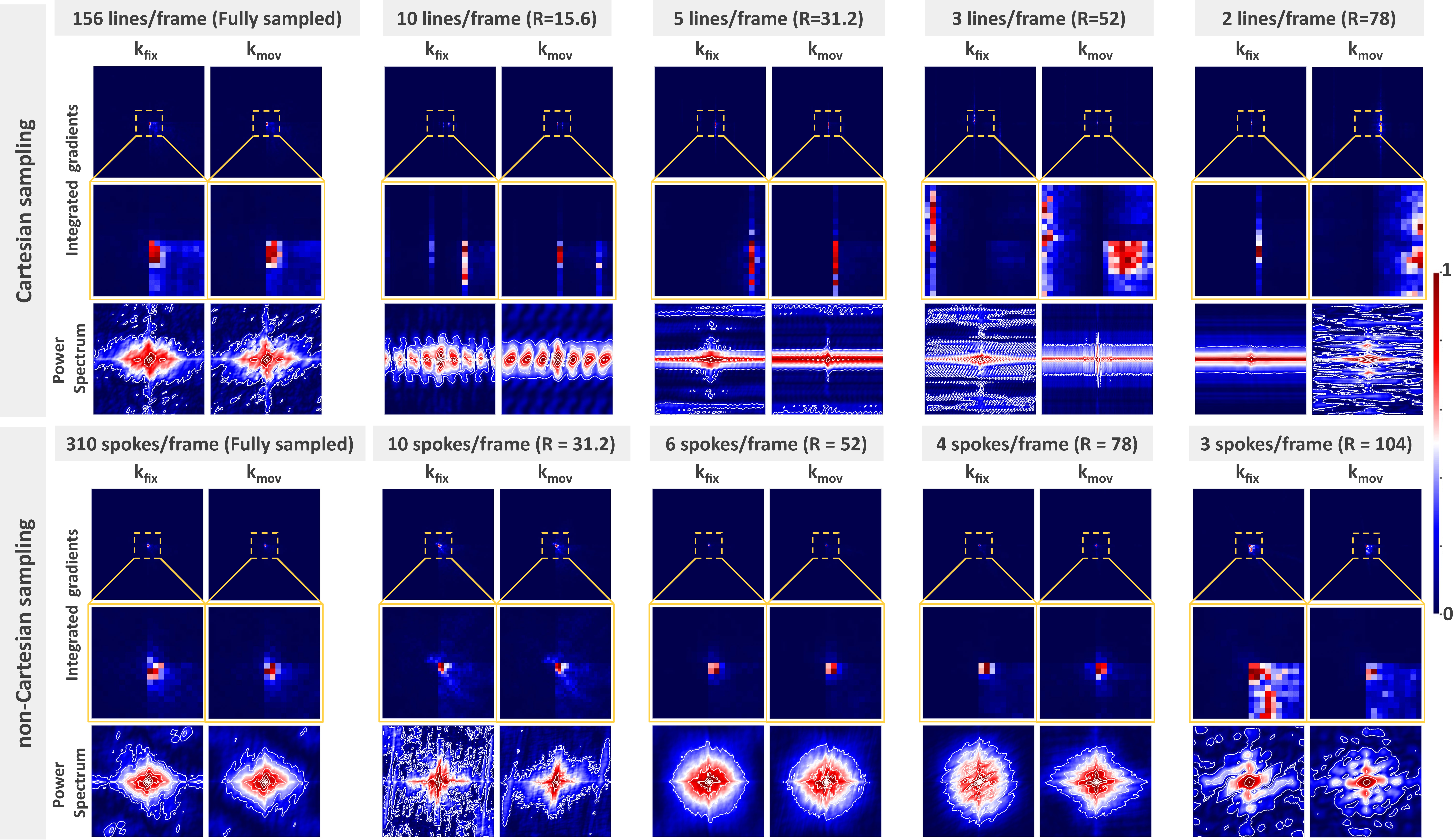}
\caption{Exemplary heatmap visualization of the Integrated Gradients for the fixed $k_{fix}$ and moving $k_{mov}$ input k-spaces and their corresponding power spectra in a representative healthy subject. Results are shown for the fully sampled case and various acceleration factors in the Cartesian VISTA and radial trajectories. The second row displays magnified views of the central region, highlighted within the yellow box. Contour lines, representing isobars in the power spectra, are outlined. The attention distributions coincided with the central region of the k-space (low-frequency range) and indicated a negative attribution towards positions in the peripheral areas (high-frequency range).}\label{ig}
\end{figure*}
\begin{figure*}[h]
\centering
\includegraphics[width=0.99\textwidth]{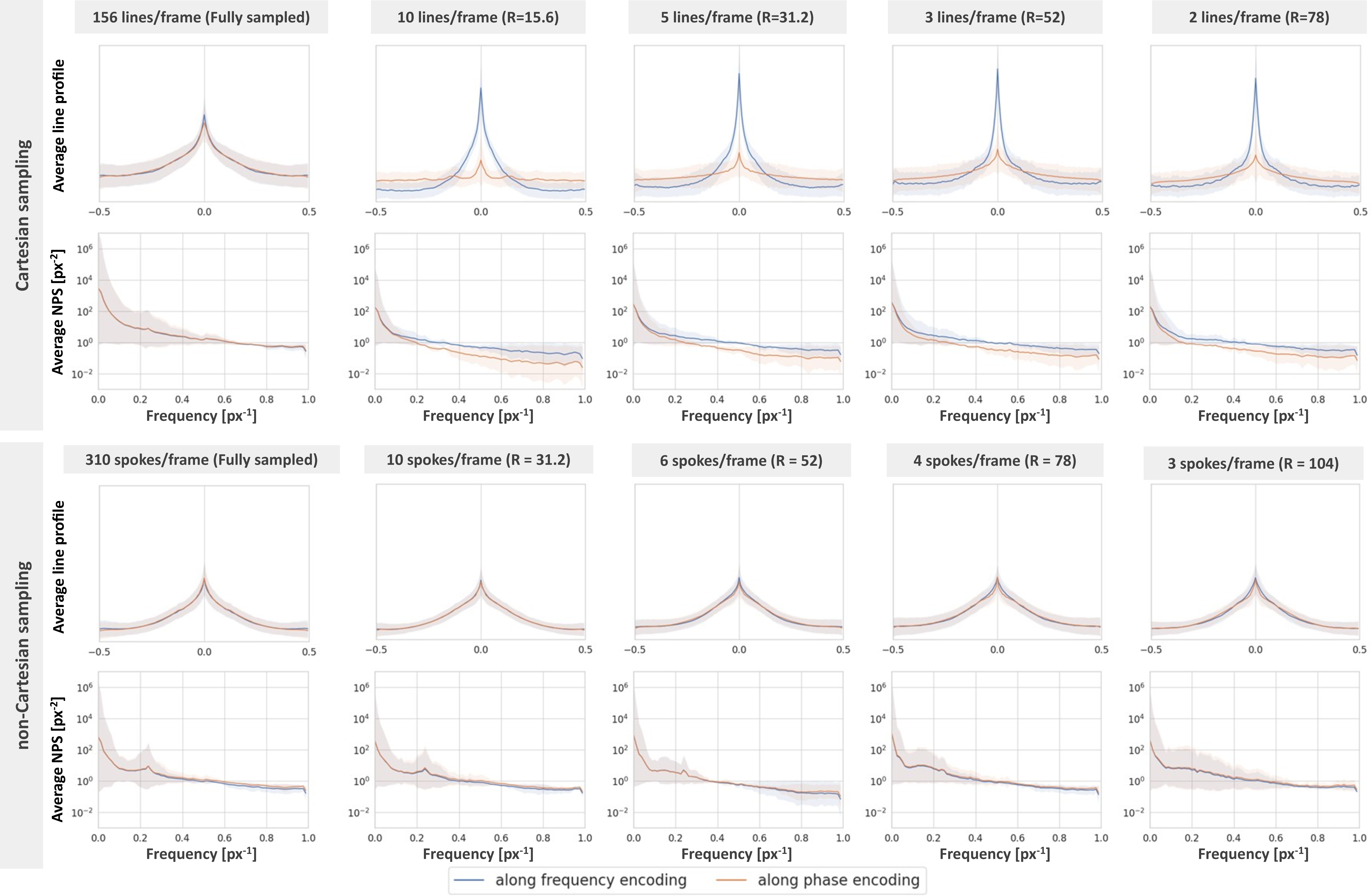}
\end{figure*}

\subsection{Ablation Studies}
We carried out multiple ablation studies with various acceleration factors to assess the contributions of the individual framework components and training strategy. The evaluation metrics are displayed in Fig.~\ref{ablations} for the fully sampled case and $R=31.2$, $R=78$, and $R=104$ acceleration rates using the radial sampling. The first set of ablation studies compared the proposed LAPANet architecture against variants without key subblocks  (w/o Global Residual Block, w/o Dilated Fusion Module, w/o Channel Integration Module, w/o Motion Attention Module) and different training loss settings (w/o data consistency loss $\mathcal{L}_{\text{K-DC}}$, w/o translational loss $\mathcal{L}_{Tphoto}$, training loss calculation with coil-combined data w/ $n_{\text{coils}}=1$). The proposed model obtained a consistent enhancement in registration performance across all studied metrics and acceleration factors compared to these versions.\\
Next, we investigated the efficacy of incorporating explicit tapered k-spaces into all processing stages (w/ tapering), rather than relying solely on the Global attention modules to implicitly perform the tapering. The additional tapering function did not enhance performance for fully sampled and intermediate investigated accelerations ($R=31.2$). Conversely, the NRMSE and HDD were significantly improved for higher accelerations with our proposed approach (NRMSE$_{\text{R=78}}$: $0.08\pm0.02$, HDD$_{\text{R=78}}$: $2.62\pm0.54$; NRMSE$_{\text{R=104}}$: $0.08\pm0.02$, HDD$_{\text{R=104}}$: $2.6\pm0.51$) in comparison to adding the tapering operation (NRMSE$_{\text{R=78}}$: $0.09\pm0.03$, HDD$_{\text{R=78}}$: $2.96\pm0.79$; NRMSE$_{\text{R=104}}$: $0.09\pm0.03$, HDD$_{\text{R=104}}$: $2.88\pm0.77$).\\ 
We also evaluated the effect of coil compression of the input k-spaces on the model performance using coil-combined ($n_{\text{coils}}=1$), 8-coil ($n_{\text{coils}}=8$), 16-coil ($n_{\text{coils}}=16$) and 24-coil ($n_{\text{coils}}=24$) resolved data. The proposed $16$-coil-resolved data yielded improved metrics compared to the more compressed data and showed no significant differences from the $24$-coil-resolved data for high accelerations. This suggests that $16$ coils offer a good trade-off between computational complexity, memory requirements, and registration performance.\\
In Supplementary Fig. A2, we presented qualitative results of the ablation studies conducted on a healthy subject for the fully sampled and accelerated cases within a radial trajectory. The depicted outcomes underscore the effectiveness of the proposed architecture and training setting in delivering consistent motion estimates. The ablated models exhibited noticeable translational shifts of the heart region, loss of details, unclear boundaries, and magnitude fluctuations across varying accelerations. 

\subsection{Model Interpretability}
To illustrate the interpretability of LAPANet, we show representative Integrated Gradient (IG) heatmaps and corresponding power spectra in Fig.~\ref{ig} for the input k-spaces in Cartesian and radial trajectories. The uncovered attention distributions aligned with the center of the k-space and showed a negative attribution for the peripheral k-space areas, in both trajectories and across different accelerations. In the case of accelerated Cartesian sampling, the power spectra were anisotropic and stretched along the phase encoding direction. In contrast, more even power spectra were observed for central k-space positions in the radial sampling.\\
The average line profiles indicated a reduction in the steepness of the slope at the edges, reflecting a decrease in the significance of high frequencies for the estimation, as depicted in Fig.~\ref{nps}. Low frequencies remained dominant as accelerations increased. The highest amplitude of the NPS sidelobes indicated an energy reduction and plateaus in the low- and intermediate-frequency regions. The decrease of the NPS values in the high-frequency region revealed the effect of low-pass filtering of high-frequency noise, substantiating that LAPANet relied more on low-frequency samples.

\section{Discussion}\label{sec4}
In this work, we introduced a novel non-rigid deep learning-based image registration in k-space, named LAPANet. A self-supervised training is adopted, eliminating the need for landmarks, additional priors, or labeling. LAPANet employs the Local-All Pass principle \cite{gilliam20163d} to decompose non-rigid motion into a series of local rigid transformations. By operating directly in k-space, the proposed network eliminates the reconstruction step and circumvents artifacts associated with high acceleration rates in the image domain. We evaluated the performance of LAPANet on cardiac MR imaging. For initial misalignments of up to $7.2$ mm, our network achieves a mean target registration error $\leq2.8$ mm in Cartesian sampling and $\leq3.3$ mm in radial sampling. Even at high accelerations ($R=78$ Cartesian, $R=104$ radial), our model demonstrates highly efficient and robust motion estimation with temporal resolutions below 5 ms. This rapid estimation capability makes our method suitable for dynamic, online, and real-time MRI applications. Qualitative and quantitative results demonstrate consistent performance across different sampling trajectories and accelerations. This adaptability ensures that sampling and temporal resolution can be customized to suit the specific needs of various applications.\\
Most existing registration methods rely on fully sampled data and, hence, require lengthy data acquisition or reconstruction beforehand \cite{balakrishnan2019voxelmorph, usman2020retrospective}. When faced with accelerated data, such methods struggle to distinguish endo- and epicardial contours in dynamic cine series. Contrarily, LAPANet eliminates the dependency on motion-resolved images and outperforms other conventional (Elastix) and deep learning-based methods (VoxelMorph and GMA-RAFT) for accelerated data. This emphasizes the feasibility of solving the registration task in k-space without being hindered by emerging streaking and blurring caused by high accelerations. Other existing works, that are more resilient to acceleration artifacts, are typically integrated into image reconstruction \cite{cordero2018three, pan2023reconstruction, yang2022end}, making it challenging to isolate the sole performance of registration. Few registration methods have been proposed for direct estimation from undersampled data \cite{huttinga2021real, kustner2021lapnet, shao2022real}. Nonetheless, these studies necessitate additional priors, such as previous scans or ground truth motion for supervised training.\\
LAPANet operates on the entire k-space to ensure a holistic understanding of the underlying data. This is especially beneficial in the case of limited remaining samples for high accelerations. As information is scattered across the entire k-space, extracting and combining low-level and high-level features become crucial for recovering intrinsic and global structures. A larger receptive field is ensured by the \textit{Dilated Fusion Module} and the \textit{Motion Attention Module} thanks to the interchangeable dilated convolutions, aggregated downsized feature maps, and attention operations. The integration of explicit tapering, a technique previously employed in LAPNet \cite{kustner2021lapnet}, to decompose non-rigid motion into local translations did not yield performance improvements in our experiments. We found that unrestricted learning of the tapered k-spaces by the Global Residual Blocks was more effective, particularly considering the limited information available in most manual tapered k-space regions at high acceleration rates. The channel-wise attention transformers in the \textit{Global Residual Module} and the \textit{Channel Integration Module} allocate attention weights to the extracted local k-space windows information according to their prevalence and mitigate the effect of the zeroed lines resulting from undersampling. By enabling the model to observe a larger receptive field and dynamically determine the most relevant features, we facilitated more efficient encoding of the widely distributed information.\\ 
The best training loss setup included the data consistency loss in k-space to emphasize missing samples due to acceleration and give global guidance during training, unlike pixel-based losses in the image domain. Moreover, we learned the translation between the input k-spaces to guide the encoder in focusing on mapping representations relevant to the registration task at earlier stages of the learning process. Omitting the coil weighting from the loss function or augmenting the coil compression resulted in substantial performance degradation. This suggests that LAPANet relies on leveraging the correlation between multiple channels of neighboring points in k-space.\\
The interpretability study demonstrated that low-frequency samples contributed more to the decision-making of LAPANet than high-frequency samples. The model attributes comparable attention weight distributions in the Cartesian and radial sampling suggesting shared functional characteristics. An optimized distribution of low-frequency and high-frequency samples can thereby be identified for overall registration improvement and can be used to create adaptive sampling patterns for motion-robust image acquisition and reconstruction.\\ 
We acknowledge certain limitations in this study. Our current approach estimates motion frame by frame, neglecting inter-frame motion smoothness and through-plane motion. By estimating the dynamics of the entire cardiac cycle, we anticipate achieving more consistent and reliable results. For that, Recurrent Neural Networks (RNNs) \cite{yang2022end}, learning bi-directional optical flows from multiple frames \cite{shi2023videoflow}, utilizing groupwise training losses \cite{pan2023reconstruction} or spatiotemporal consistency regularization \cite{asif2013motion} can be utilized. Furthermore, our training loss function depends on fully sampled data. While cine is a standard acquisition in clinical practice, and LAPANet operates without fully sampled data post-training, this requirement remains suboptimal since collecting fully sampled data is still necessary to train and fine-tune new models.\\ 
This study focuses on cardiac motion estimation, laying the groundwork for future investigations of other motion types. Future investigations on larger cohorts, prospectively acquired data, other imaging sequences, and motion patterns are warranted to verify the generalizability of LAPANet. Furthermore, we plan to extend the proposed framework towards a foundational model for deep learning-based non-rigid motion estimation, thereby broadening its scope of application.\\
The presented registration technique is useful for a wide range of cardiac imaging tasks. Unlike ECG-synchronized MRI, our method can represent beat-to-beat motion variations without being affected by physiologic influences such as respiration, blood pressure, heart rate, exercise, or medication. This eliminates the need for breathing protocols and improves patient compliance. Moreover, our method has the potential to expand the diagnostic capabilities of cardiac MRI by providing functional access to individual cardiac cycles. Key applications include real-time assessment and monitoring of myocardial function, cardiac valve dynamics, turbulent blood flow, real-time visualization of rapid physiological processes, and MRI-guided catheterization and interventions \cite{zhang2014real}.

\section{Conclusion}\label{sec5}
In conclusion, we introduced a new deep learning-based method for non-rigid registration in k-space and validated its efficacy for estimating motion in highly accelerated (sub-heartbeat) cardiac MR imaging. In a multi-resolution processing pipeline, we model the mapping between the fixed and moving images relying on interleaved attention modules, parallel dilated convolution operations, and the reuse and aggregation of multi-scale feature representations. Unlike image-based methods, LAPANet demonstrated sustained performance across various trajectories and acceleration factors, adapting to variable temporal resolutions. The achieved proficiency at heightened accelerations makes our method particularly valuable for enhancing the diagnostic capabilities of motion imaging applications and revealing previously undiscovered motion dynamics that can ultimately improve patient outcomes. 

\section*{Acknowledgments}
The work was supported by the Deutsche Forschungsgemeinschaft (DFG, German Research Foundation) under Germany’s Excellence Strategy – EXC 2064/1 – Project number 390727645.

\bibliographystyle{unsrt}  
\bibliography{references}

\end{document}